\documentclass[10pt]{IEEEtran}

\usepackage{tikz}
\usepackage{amsmath}
\usepackage{xspace}
\usepackage{enumitem}
\usepackage{amsfonts}
\usepackage{graphicx}
\usepackage{tikz}
\usepackage{amsmath}
\usepackage{times}
\usepackage{subfigure}
\usepackage{caption} 
\usepackage{algorithm}
\usepackage{algorithmic}
\usepackage{tabularx}
\usepackage{xspace}
\usepackage{booktabs}
\usepackage{multirow}
\usepackage{enumitem}
\usepackage{makecell}
\setlist{nosep}
\usepackage{url}
\usepackage{orcidlink}
\usepackage{amsmath,amssymb,amsfonts}
\usepackage{textcomp}
\usepackage{xcolor}
\usepackage{xspace}
\usepackage{float}
\usepackage{comment}
\usepackage{listings}

\usepackage{amsmath}
\usepackage{amssymb}
\usepackage{mathtools}
\usepackage{amsthm}

\hypersetup{
  colorlinks=false,
  linkbordercolor=white,
 urlbordercolor=white,
pdfborder={0 0 0}
}

\definecolor{codegreen}{rgb}{0,0.6,0}
\definecolor{codegray}{rgb}{0.5,0.5,0.5}
\definecolor{codepurple}{rgb}{0.58,0,0.82}

\definecolor{backcolour}{rgb}{1,1,1}

\lstdefinestyle{mystyle}{
    backgroundcolor=\color{backcolour},   
    commentstyle=\color{codegreen},
    keywordstyle=\color{magenta},
    numberstyle=\tiny\color{codegray},
    stringstyle=\color{codepurple},
    basicstyle=\ttfamily\footnotesize,
    breakatwhitespace=true,         
    breaklines=true,                 
    captionpos=b,                    
    keepspaces=true,                 
    numbers=left,                    
    numbersep=5pt,                  
    showspaces=false,                
    showstringspaces=false,
    showtabs=true,                  
    tabsize=4,
    frame=trbl,
    rulecolor=\color{black},
    linewidth=0.98\linewidth,
     belowskip=-0.1in
}

\newcommand{\system}{{\textsc{\small{Dalc-CT}}}\xspace}

\begin{document}

\title{DALC-CT: Dynamic Analysis of Low-Level Code Traces for Constant-Time Verification}

\author{
Nges Brian Njungle \orcidlink{0009-0006-3393-6851}, 
Edwin P. Kayang \orcidlink{0009-0003-7158-5608}, 
 Mishel J. Paul  \orcidlink{0009-0005-1002-2515}, 
and Michel A. Kinsy  \orcidlink{0000-0002-1432-6939} \\
STAM Center, Ira A. Fulton Schools of Engineering\\
Arizona State University, Tempe, AZ 85281, USA\\
Emails: nnjungle@asu.edu, ekayang@asu.edu, mpaul16@asu.edu, mkinsy@asu.edu
}

\maketitle

Timing side-channel attacks exploit variations in program execution time to recover sensitive information. 
Cryptographic implementations are especially vulnerable to these attacks, since even small timing differences in operations such as modular exponentiation or key comparisons can be exploited to extract highly sensitive information, such as secret keys. 
To mitigate this threat, implementations of programs that handle sensitive information are often expected to adhere to constant-time principles, ensuring that execution behavior does not depend on secret inputs. 
However, validating the constant-time property of programs remains a major challenge in cryptography development. 
Formal method approaches to verify constant-time implementations rely on abstractions that often fail to capture real execution behavior, while timing-based measurement techniques are highly sensitive to noise from other programs and even hardware environments. 

In this work, we propose a novel approach for verifying constant-time programs based on dynamic analysis of low-level execution traces. 
Our method measures instruction sequences across multiple input values for any given binary and targeted function.
Any variations in the instruction mix distribution for any given pair of traces indicate a deviation from the constant-time principle and behavior. 
We developed an open-source tool called \system, for the constant-time verification of programs using this approach. 
We evaluated it on a set of well-known constant-time and non-constant-time examples, achieving a perfect detection of issues. 
Our results demonstrate that analyzing the logical execution of programs via instruction trace comparisons provides a lightweight and reliable way to verify the constant-time property of programs.


\section{Introduction}

Computer vulnerabilities are weaknesses or flaws in software, hardware, or system processes that can be exploited to compromise the confidentiality, integrity, or availability of information systems \cite{igure2008taxonomies}. 
These vulnerabilities come in many forms, including software bugs, design flaws, misconfigurations, insecure cryptographic protocols, weak authentication mechanisms, social engineering risks, and hardware-level exploits \cite{aslan2023comprehensive}. 
They often arise from poor coding practices, inadequate testing, outdated software components, or even human errors.
They provide malicious actors with opportunities to gain unauthorized access, disrupt operations, steal data, or escalate privileges in computer systems \cite{pfleeger2012analyzing}. 
Among the more subtle vulnerabilities are side-channel vulnerabilities, which exploit indirect information leaks from a system’s physical or logical behavior rather than direct flaws in its code or logic \cite{standaert2009introduction}. 
For example, attackers can analyze timing variations, power consumption, electromagnetic emissions, or even cache access patterns to infer sensitive data such as encryption keys. 
Side-channel attacks are particularly concerning because they bypass traditional security boundaries, often exploiting low-level system characteristics that are difficult to detect or patch through conventional software updates \cite{palihawadanasilent}.

Timing side-channel attacks are a broad and well-known subclass of attacks that exploit side-channel vulnerabilities by analyzing and exploiting the variations in the time it takes a system to perform certain operations or tasks \cite{reparaz2017dude}. 
Even small differences in processing time, such as those occurring during password validation or cryptographic computations, can leak valuable sensitive information. 
By repeatedly measuring and analyzing these time differences, attackers can gradually deduce sensitive data, including secret keys or authentication tokens. 
Timing attacks are particularly dangerous because they do not rely on exploiting software bugs directly but instead leverage observable performance characteristics that may appear harmless yet expose critical weaknesses in critical systems \cite{zaidenberg2015timing}.

Paul Kocher published the first set of timing side-channel attacks in 1996, demonstrating that secret keys used in cryptographic algorithms such as RSA, Diffie-Hellman, and other cryptographic systems could be recovered by carefully measuring the time required to perform the modular exponentiation module of these algorithms\cite{kocher1996timing}. 
Since then, timing-based exploits have been shown to affect a wide range of critical real-world systems. 
For example, Brumley et al. demonstrated remote key recovery attacks against OpenSSL’s implementation of RSA decryption \cite{brumley2005remote}. 
Canvel et al. \cite{canvel2003password} revealed timing leaks in SSL/TLS implementations that enabled attackers to recover session keys, while Lucky et al. \cite{al2013lucky} introduced timing-based padding-oracle attacks capable of extracting private data from SSL/TLS connections. 
Widely known timing side-channel microarchitectural attacks such as Spectre and Meltdown have also illustrated how simple it is to leverage the minimal execution time differences at the hardware level to leak sensitive information even from protected memory regions\cite{kocher2020spectre,lipp2018meltdown}. 

One of the most effective defenses against timing attacks is the use of constant-time code. 
Here, programs are implemented in such a way that their execution time does not depend on secret values ~\cite{jancar2022they}. 
In constant-time programming, algorithms are carefully designed to avoid data-dependent branches, memory accesses, and arithmetic operations that could introduce measurable timing variations. 
A common example is password comparison, where instead of terminating as soon as a mismatch is detected, all characters are checked in sequence so that the total execution time remains constant. 
Similarly, in cryptographic libraries, arithmetic routines are also implemented in a way that ensures every operation takes the same number of cycles regardless of the input, preventing attackers from observing key-dependent variations.
Despite its effectiveness, implementing constant-time programs is extremely difficult. 
It requires a deep understanding of low-level programming, instruction-level timing, compiler optimizations, and microarchitectural behavior \cite{cauligi2017fact}. 
Developers must often abandon conventional coding practices and instead adopt specialized techniques to control every aspect of the execution of their code \cite{almeida2016verifying}. 
Even then, constant-time behavior is not guaranteed, since compilation and optimization levels may reintroduce timing variations in the program's executable instructions.

This raises the important question of how to verify whether a program's implementation is truly constant-time. 
Dudect proposed a practical approach that combines leakage detection with statistical analysis by measuring the execution time of code across multiple inputs on a given platform to determine the non-constant-time property~\cite{reparaz2017dude}. While effective in certain scenarios, this approach is highly sensitive to noise which is often present in  modern computer systems. 
The heterogeneous nature of modern systems allow for the execution of multiple processes and concurrent background tasks. As a result, it is extremely difficult to guarantee consistent and accurate measurements of execution time, which means the method can only detect constant-time violations that produce large timing variances.
Furthermore, Almeida et al. \cite{almeida2016verifying} developed ct-verif and provided a formal foundation for constant-time programming policies, using a leakage model that captures different notions of constant-time behavior such as path-based, address-based, and operand-based policies. 
Several other tools, including Binsec/Rel \cite{daniel2020binsec}, Blazzy et al \cite{blazy2019verifying},  and CTgrind  \cite{langley2010ctgrind}, have explored validation methods based on formal verification, LLVM~\cite{lattner2004llvm} bitcode, and  binary analysis. 
While useful, these tools often rely on abstractions or simulations that cannot fully capture the real-world hardware execution of programs.

In this work, we introduce a novel approach to constant-time code verification that leverages the dynamic analysis of low-level execution traces. 
Rather than analyzing source code or relying on simulated execution, our method measures the sequence of instructions executed across different inputs  and checks for variations in the instruction traces. 
By observing the program’s logical execution paths at the instruction level under diverse execution conditions, our approach provides a precise assessment of whether an implementation maintains constant-time execution behavior independent of it's inputs and system software. 
This approach avoids dependence on external factors available in the system and instead captures the internal execution characteristics of the program, enabling a very reliable way for detecting constant-time violations.

Based on this approach, we implement an open-source tool called \system. 
Given a program as a binary, \system accepts a target function and the secret input variable to verify. 
The program execution traces are analyzed repeatedly across varying inputs of the secret input variable. 
During execution, \system records the sequence of instructions for the target function and classifies them into thirteen categories according to instruction types and data types. 
Any variation in instruction mix counts across input values is treated as a clear indication of a deviation from constant-time behavior.
While this method does not directly capture some microarchitectural effects such as caching and speculative execution, it offers a practical and concrete way to detect deviations from constant-time property at the lowest level of execution logic, bridging the gap between theoretical verification and real-world constant-time code behavior in programs.
Our main contributions in this work include:  
\begin{itemize}  
    \item We introduce a new approach for the verification of constant-time programs based on dynamic analysis of low-level execution traces. By analyzing instruction mix counts, this approach provides a concrete and reliable way of evaluating the logical execution behavior of programs at the lowest level of logic, independent of system software, compilers, optimization levels, and hardware implementation.  
    \item We develop an open-source tool called \system based on this approach. \system is evaluated on a wide range of well-known constant-time and non-constant-time examples. It detects non-constant-time behavior in all non-constant-time cases using just a few rounds.  
\end{itemize}

\section{Related Works}
Tools for verifying constant-time code are generally classified into three categories: runtime statistical tests, dynamic instrumentation, and formal analysis-based approaches \cite{jancar2022they}.
Each category tackles the problem from a different perspective and offers distinct advantages and limitations.

Runtime Statistical Tests tools detect timing leaks by repeatedly executing a program and measuring its runtime under different inputs. 
They then apply statistical methods to determine whether variations in execution time correlate with secret data. While simple and lightweight, this approach is inherently limited because runtime measurements are highly susceptible to noise from the operating system, and other concurrent processes. 
For instance, Dudect \cite{reparaz2017dude} collects a large number of execution-time samples and applies Welch’s t-test to identify statistically significant differences in timing distributions, flagging potential non-constant-time programs. RTLF~\cite{dunsche2024great}, another tool within the category, uses a technique based on bounded type-1 errors to try to minimize false positives.

Dynamic Instrumentation tools analyze how data flows during execution by instrumenting the program, either for concrete or symbolic values using multiple runs. 
These tools often operate on intermediate representations like LLVM~\cite{lattner2004llvm} bitcode, which makes them effective at detecting secret-dependent branches. 
For instance, CTgrind \cite{langley2010ctgrind} extends Valgrind~\cite{nethercote2007valgrind} to check for runtime secret-dependent branches via memory accesses, while Binsec/Rel \cite{daniel2020binsec} used symbolic execution and relational verification to determine whether different secret inputs lead to identical control-flow and memory-accesses.
Although powerful and provide the most reliable way for constant-time analysis, current tools still rely on some level of abstraction thus, do not fully capture the logical  behavior of programs on hardware.

Formal-analysis based approaches use mathematical reasoning to guarantee that implementations adhere to constant-time principles. These methods typically rely on source code, intermediate representations, or formal models of hardware, which makes them theoretically strong but difficult to scale and apply in practice. Almeida et al. \cite{almeida2013formal} proposed a type-system that enforces path, address, and operand-based constant-time policies. z
Building on this foundation, ct-verif \cite{almeida2016verifying} provides automated verification for C implementations of cryptographic algorithms using symbolic execution and verification conditions. 
FlowTracker \cite{rodrigues2016sparse} applies static information-flow analysis to detect whether secret data can influence control flow or memory access. 
While these tools provide strong theoretical guarantees, they require complex modeling and often struggle with completeness, making them challenging to use.

Although these approaches have significantly advanced the verification of constant-time code, each category has fundamental limitations. 
Runtime statistical tests are practical but offer only weak guarantees due to susceptibility to measurement noise. 
Dynamic instrumentation improves observability but usually relies on abstract program representations that often diverge from actual hardware execution. 
Formal-analysis tools provide strong guarantees but are difficult to scale and use, as they depend on complex models that are often incomplete. 
These limitations point to the need for complementary techniques that directly capture the constant-time behavior of programs. 
This work addresses this gap by introducing a method based on dynamic analysis of low-level execution traces. 
Rather than relying on external timing measurements or abstract models, our approach records and analyzes the actual instructions executed by compiled binaries under different input values of the secret input. 
By comparing instruction mix counts across different inputs, we directly detect whether secret input values cause variations in a program's execution behavior. 
This approach combines the concreteness of observing program logical behavior on hardware with the practicality of an automated workflow, offering a precise, lightweight, and accessible approach to verify the constant-time property of programs.

\section{Background}
When a program is written in a high-level language like C or C++, it undergoes a series of transformations before it can be executed on hardware. 
First, the source code is processed by the compiler, which lexes and parses the program, translates it into intermediate representations, and applies optimizations before generating low-level assembly~\cite{alfred2007compilers}. 
During this stage, compiler optimizations such as constant folding, loop unrolling, or dead-code elimination may significantly alter how the code executes, sometimes in ways that are not evident from the source code. For reference, LLVM exposes over a hundred optimization flags. Different permutations of these produce varying effects on the compiled code~\cite{ashouri2017micomp}.
Next, an assembler translates the optimized assembly code into machine code.
Following this, the linker combines the assembled program with external libraries and resolves references to functions and variables, producing an executable binary~\cite{presser1972linkers}. 
During execution, this binary is loaded into memory by the operating system, which allocates resources and sets up the runtime environment.
Finally, the processor executes the program as a sequence of low-level instructions~\cite{agarwal1990april}.
These layered transformations from source code to execution make it challenging to fully understand any program’s logical behavior, highlighting the importance of analysis at the lowest logical level of a program. 

Because the path from source code to execution involves multiple transformations, constant-time properties asserted at the source level or through abstraction layers may not hold once the program is compiled and executed. 
As a result, analyzing constant-time behavior at these levels risks overlooking minute but exploitable timing differences that may occur in the executable program. 
On the other hand, instruction-level analysis provides a more reliable foundation, as it directly examines the concrete execution paths of the program on hardware. 
Verifying constant-time properties at this level offers stronger guarantees that a program behaves consistently across all secret inputs.

Different instructions exhibit different execution behaviors. 
On \textit{x86} processors for example, the execution time of memory instructions such as loads or stores depends on whether the data is already in cache or must be fetched from main memory. 
Arithmetic Logic Unit (ALU) instructions perform operations such as addition, subtraction, bitwise manipulation, multiplication, and division. Although these instructions are often considered “fast”, their execution latencies can differ substantially. Simple operations like addition, subtraction, and bitwise logic typically complete in a single cycle, whereas more complex operations such as multiplication and division incur significantly higher latencies, often spanning multiple cycles. These differences are largely determined by the underlying hardware implementation and available execution units.
Control-flow instructions, including conditional and unconditional branches, can also impact execution time. Their cost is highly dependent on microarchitectural factors such as branch prediction accuracy and pipeline depth, with mispredicted branches causing pipeline flushes and introducing nontrivial timing penalties.

Moreover, the data types and operand sizes used in a program, such as integers, floating-points, or vectors, also influence both the number of instructions required and their execution latencies. 
As a result, even programs that are functionally equivalent can exhibit measurable differences in execution time depending on the mix of instructions and the data-flow, making precise constant-time verification very extremely challenging.

A constant-time program is one whose execution behavior does not depend on secret input values~\cite{bernstein2013mcbits}. 
In practice, this means that the sequence of instructions executed, the data-flow, and the timing of operations remain identical regardless of the values of a sensitive input. 
The key requirement is that no observable property, such as execution time, control-flow decisions, or memory access patterns, varies as a function of secret information. 
By analyzing a program's instruction traces over different secret input values, it is expected that a constant-time program will exhibit the exact same logical behavior for every possible value of the secret input. 
In other words, the instruction traces should be identical across all secret inputs, ensuring that no exploitable variations in execution time exist.
These identical instruction sequences across all secret inputs provides stability in how the hardware executes the program ensuring consistency in control flow, memory access patterns, and execution time thus the only reliable way of verifying constant-time behavior.
\section{Formal Construction of Constant-Time Property}

In this section, we provide a formal framework for verifying constant-time behavior in programs using low-level execution traces. This approach captures instruction-level execution properties independent of compiler optimizations, system software, or hardware implementation, offering a concrete and implementation-grounded definition of constant-time behavior.

\subsection{Execution Traces and Instruction Mix}

Let $P$ be a program and $f \subseteq P$ a target function. The input space of the program is partitioned into public inputs $\mathbf{p} \in \mathcal{P}$ and secret inputs $\mathbf{s} \in \mathcal{S}$. We model the execution of function $f$ on input $(\mathbf{p}, \mathbf{s})$ as a sequence of instructions:
\[
\mathcal{E}(f, \mathbf{p}, \mathbf{s}) \rightarrow \tau
\]
where $\tau = (i_1, i_2, \dots, i_k)$ is the trace of instructions executed during the evaluation of $f$.  

To analyze execution behavior, we classify instructions into a finite set of classes:
\[
\mathcal{C} = \{c_1, c_2, \dots, c_m\},
\]
covering arithmetic, memory, and control-flow operations. A classification function 
\[
\kappa : \text{Instr} \rightarrow \mathcal{C}
\]
maps each instruction to its class. For a given trace $\tau$, we construct the \emph{instruction mix vector}:
\[
\Phi(\tau) =
\begin{bmatrix}
\phi_1 \\ \phi_2 \\ \vdots \\ \phi_m
\end{bmatrix}
\in \mathbb{N}^{m \times 1}, \quad
\phi_j = \big|\{ i \in \tau \mid \kappa(i) = c_j \}\big|
\]
where each entry $\phi_j$ counts the occurrences of instructions in class $c_j$. This vector representation allows for precise mathematical comparison of execution behavior across different secret inputs.

\subsection{Trace Equivalence and Constant-Time Property}

We define an equivalence relation between traces based on their instruction mix:
\[
\tau_1 \equiv_{\Phi} \tau_2 \;\;\triangleq\;\; \Phi(\tau_1) = \Phi(\tau_2)
\]
This equivalence abstracts away instruction ordering but preserves the quantitative characteristics of execution at the category level.  

A function $f$ satisfies trace-based constant-time behavior with respect to secret inputs if for all $\mathbf{s}_1, \mathbf{s}_2 \in \mathcal{S}$:
\[
\Phi\big(\mathcal{E}(f, \mathbf{p}, \mathbf{s}_1)\big)
=
\Phi\big(\mathcal{E}(f, \mathbf{p}, \mathbf{s}_2)\big)
\]
Equivalently, using the Manhattan norm on the vector space $\mathbb{N}^{m \times 1}$:
\begin{align}
f \text{ is constant-time} 
& \iff \forall \mathbf{s}_1, \mathbf{s}_2 \in \mathcal{S}, \nonumber \\
& \quad \|\Phi(\mathcal{E}(f, \mathbf{p}, \mathbf{s}_1)) 
- \Phi(\mathcal{E}(f, \mathbf{p}, \mathbf{s}_2))\|_1 = 0
\end{align}

where the norm sums the differences across all classes. Any non-zero difference indicates secret-dependent variation in execution.

\subsection{Practical Approximation and Verification}

Exhaustive verification over the entire secret input space $\mathcal{S}$ is generally infeasible. To make verification practical, we consider a finite sampling set $\hat{\mathcal{S}} \subseteq \mathcal{S}$:
\[
\forall \mathbf{s}_i, \mathbf{s}_j \in \hat{\mathcal{S}}, \quad
\Phi(\mathcal{E}(f, \mathbf{p}, \mathbf{s}_i)) = \Phi(\mathcal{E}(f, \mathbf{p}, \mathbf{s}_j))
\]
Any counterexample observed within $\hat{\mathcal{S}}$ constitutes evidence of a constant-time violation. This empirical approach enables reliable assessment of program behavior while remaining tractable for real-world programs.

This framework formalizes constant-time verification as a structural invariant over instruction traces, providing a robust method for detecting secret-dependent execution differences. By representing traces as instruction mix vectors and using equivalence relations or norms for comparison, our approach bridges the gap between theoretical constant-time definitions and practical, binary-level program verification. It captures the logical execution characteristics of a program, avoiding reliance on high-level source code or timing measurements, and ensures that any observed deviations are attributable solely to secret-dependent behavior.

\section{System Design and Implementation}

\begin{figure*}[http]
    \centering
    \includegraphics[width=\linewidth]{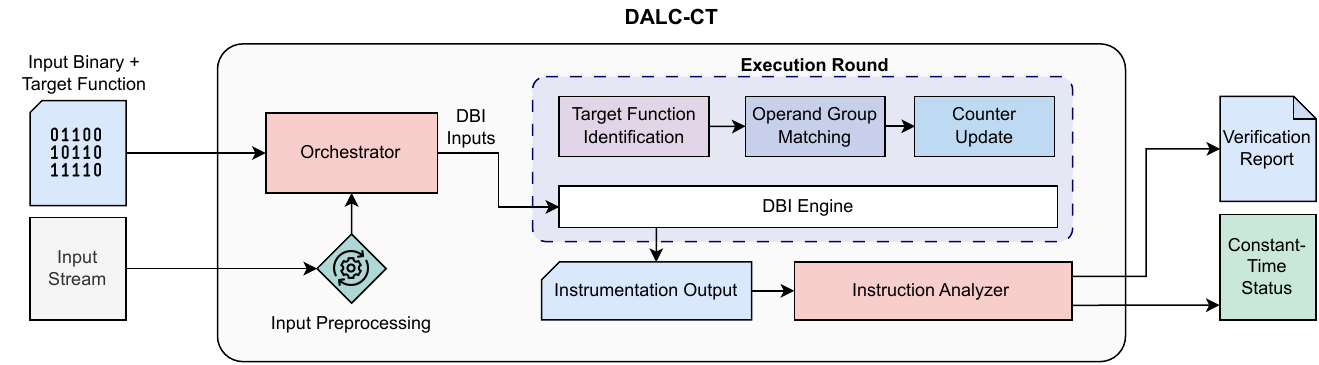}
    \caption{Architecture of \system. The framework takes a program binary, secret input values, and the target function, then uses a DBI engine to instrument execution rounds. The orchestrator manages DBI inputs, instrumentation while the instruction analyzer evaluates operand groups and instruction behavior to determine the constant-time status and generate a verification report.}
    \label{fig:system_fig}
    \vspace{-0.2in}
\end{figure*}

\system is a constant-time verification tool that analyzes a program’s low-level instruction traces under different secret input values and identifies potential violations by detecting variations in execution behavior, thereby assessing it's constant-time property. Figure~\ref{fig:system_fig} illustrates the architecture of \system, showing how it receives inputs and leverages a Dynamic Binary Instrumentation (DBI) engine to analyze program traces.

It accepts a compiled binary together with the target function name and secret input values as inputs. 
Using a DBI engine, \system traces the execution path of the specified function under multiple input values.
The DBI engine records detailed instruction-level traces, which are then processed by the instruction mix logging routine. 
This routine organizes the collected traces, counting the number of instructions for every single instruction in the trace and creating a structured output for further analysis. 
The instruction analyzer examines these logs to detect instruction types, and classify them into different classes. It then determine variations in instruction counts from these classes across different secret input values. 
By identifying whether execution paths or instruction mixes differ, \system determines whether the program upholds the constant-time property.
\system outputs a verification report summarizing the instruction mix counts, execution behavior, and the constant-time verification status.

To analyze instruction traces, we used Valgrind as our DBI engine~\cite{nethercote2007valgrind}.
Valgrind is a very popular instrumentation framework for building dynamic analysis tools~\cite{nethercote2007valgrind}.
We chose it because of  the stability and reliability of it's DBI engine which has already been used to build well-known dynamic analysis tools.  
We implemented a valgrind routine that identifies the target function, carryout instructions grouping, and count the number of instructions for every instruction type for every execution trace of the input binary. 
Multiple instances of each binary are run through this routine using an input stream of varying input values for the secret input, and the resulting instruction sequences are collected. 
The instrumentation output is passed to our instruction analyzer for analysis and verification. 
Instructions are classified into thirteen categories based on instruction type and data type. 
These categories include memory instructions, ALU instructions, and control-flow instructions, while the data types considered are integers, floating-points, and vectors. 

We further partition ALU instructions into two categories based on their execution latency measured in CPU cycles. Instructions that typically complete within a small, bounded number of cycles are classified as light ALU instructions, whereas instructions that incur significantly higher latencies are classified as heavy ALU instructions.
Light ALU instructions include bitwise operations, addition, and subtraction, which are commonly implemented as single-cycle or low-latency operations. In contrast, heavy ALU instructions include multiplication, division, and remainder operations, which are generally multi-cycle.
This classification is introduced to improve the fidelity of our analysis by accounting for heterogeneous ALU instruction latencies. In particular, it prevents equating counts of single-cycle instructions with counts of multi-cycle instructions, a simplification that may  obscure substantial latency differences and lead to misleading conclusions when compared against actual hardware execution behavior.

Since data types also affect the performance of memory and ALU instructions, our classification results in a total of thirteen instruction classes, capturing all relevant instruction and data type variations.
We assumed that the operand sizes does not affect execution time of a program. 
This assumption holds for \textit{x86} and ARM architectures, where word-sized operations execute in a fixed number of cycles. 
This simplifications allows \system to maintain a manageable classification of instructions without an exponential increase in categories. 
Table~\ref{tab:instruction_classes} shows all the instruction classes of \system. 
This classification is essential for enabling scalable analysis of instruction-level behavior across architectures. By grouping instructions into classes, \system can easily reason about execution latency and potential variability without tracking each individual instruction encoding. This abstraction is particularly important given the vast diversity of instructions on the different Instruction Set Architectures (ISAs). For example, the x86 ISA alone exposes well over 3,000 distinct instruction variants when accounting for operand sizes, addressing modes, prefixes, and vector extensions. 
Analyzing each variant independently would be impractical thus, our instruction classes provide a compact yet expressive representation that captures all performance-relevant distinctions.

For each execution instance, \system records the counts for all instruction classes. 
If the count values for any instruction class vary across any pair of secret input values, the program is determined to violate the constant-time property. 
This approach ensures that even subtle variations in instruction execution caused by secret-dependent logic or compiler optimizations are reliably detected.
Because it operates directly on instruction traces and relies on a limited set of well-defined instruction classes, \system offers a lightweight and practical solution for constant-time verification. Moreover, \system is independent of system software, compilers, compilation optimizations, and hardware architecture, making it suitable for real-world detection of constant-time violations in program logic.
For specialized architectures or extended applications, \system instruction classification can be easily extended to add new classes that better reflect the given application scenario.

\subsection{Granularity of Instruction Categories}

Our current categorization strategy adopts a deliberately small number of instruction classes to balance detection capability with interpretability and efficiency. While the number of classes is limited, no information is discarded. Fine-grained details such as specific instruction variants and operand forms (e.g., register--register, register--memory, register--immediate) remain available and can be incorporated when deeper analysis is required.

The motivation for grouping instructions into a small set of categories stems from the complexity of modern processors. Contemporary CPU architectures expose thousands of instructions, making per-instruction analysis both computationally expensive and conceptually unwieldy. Moreover, although instructions differ semantically, many exhibit similar latency characteristics, while others vary depending on microarchitectural features and execution context. Our approach addresses this by grouping instructions into classes that reflect broadly similar timing behavior, enabling efficient reasoning about constant-time properties without incurring prohibitive analysis overhead.
Importantly, this coarse-grained design does not compromise effectiveness. In fact, fewer classes directly translate to faster instrumentation and simpler downstream analysis. In our evaluation, the proposed classification was sufficient to capture all non-constant-time behaviors across the tested examples with minimal effort. Using a lightweight havoc-based fuzzing strategy, no benchmark required more than five runs to expose timing variations.

While our current system prioritizes minimalism for efficiency and scalability, the granularity of categorization in \system can be increased as needed. A more fine-grained classification can provide additional diagnostic insight, allowing developers to pinpoint the exact instructions responsible for constant-time violations and to guide targeted remediation efforts. This flexibility enables \system to serve both as a lightweight detection tool and as a foundation for more detailed performance and security analysis.

\begin{table}[htbp]
\centering
\caption{Instruction Classification by Type, Class, and Data Type proposed and used in \system}
\label{tab:instruction_classes}
\begin{tabularx}{\linewidth}{|l|l|X|}
\hline
\textbf{Instruction Type} & \textbf{Instruction Class} & \textbf{Data Type} \\ \hline
\multirow{6}{*}{Memory} & \multirow{3}{*}{Load} & Integer \\ \cline{3-3} 
 & & Floating-point \\ \cline{3-3} 
 & & Vector \\ \cline{2-3} 
 & \multirow{3}{*}{Store} & Integer \\ \cline{3-3} 
 & & Floating-point \\ \cline{3-3} 
 & & Vector \\ \hline
\multirow{6}{*}{ALU} & \multirow{3}{*}{Light} & Integer \\ \cline{3-3} 
 & & Floating-point \\ \cline{3-3} 
 & & Vector \\ \cline{2-3} 
 & \multirow{3}{*}{Heavy} & Integer \\ \cline{3-3} 
 & & Floating-point \\ \cline{3-3} 
 & & Vector \\ \hline
Control-Flow & Conditional / Unconditional & -- \\ \hline
\end{tabularx}
\end{table}
\section{Evaluation}
\label{sec:evaluation}
To evaluate \system, we tested twenty-two  well-known constant-time and non-constant-time examples and analyzed the results. We used all examples provided in the GitHub repository ct-tools~\cite{cttools-exams}, which combines the examples presented by~\cite{2022-sp-jancar} and~\cite{2024-usenix-jancar}. Our verdicts were consistent with the expected results from the benchmarks. In addition, we included widely used non-constant-time and constant-time implementations such as the AES rijndaelKeySetupEnc function, string subcopy, modular exponentiation, and password checking.
In this paper, we present two detailed case studies for the password checking and modular exponentiation examples. We use these case-studies to illustrate both constant-time and non-constant-time variants, highlighting how each implementation either respects or violates the constant-time property. We selected these examples because they are complete, highly relevant for security applications, and simple enough to be easily demonstrated.
We present results from \system on multiple binary instances of these programs, compiled with GCC and Clang at optimization levels \textbf{-O0}, \textbf{-O1}, and \textbf{-O2}. 
The evaluation demonstrates consistent characterization of a program’s constant-time behavior across compilers and optimization settings. 
Overall, this assessment shows that \system reliably detects both constant-time and non-constant-time behavior across diverse programs, independent of system software, compiler or optimization level, thereby highlighting the effectiveness and robustness of our approach.

For the case studies, we present a set of handpicked inputs for illustrative purposes. In \system, however, we implement a havoc-based fuzzer that automatically generates randomized inputs to explore the program’s execution traces. This process is executed iteratively to collect instruction traces across diverse program behaviors. 

\subsection{Password Checking  Case Study}
Password verification routines are a well-known source of timing side channels when implemented in a naive manner. 
Consider the function in Listing \ref{listing:password}, which performs a straightforward character-by-character comparison between the secret password \texttt{pw} and the user-provided input \texttt{in}.
While simple and functionally correct, this implementation is not constant-time. 
The loop exits as soon as a mismatch is encountered, meaning that the number of executed instructions depends on how many characters of the input match the secret prefix. 
As a result, inputs that share a longer prefix with the secret password trigger longer execution traces, creating observable timing differences in different inputs matched against the secret input. 

\lstset{style=mystyle}
\begin{lstlisting}[language=C, float, caption=Password Checking Non-constant-time Example, label={listing:password}]
    int password_checker(char *pw, char *in, int len) {
        for (int i = 0; i < len; i++) {
            if (pw[i] != in[i]) {
                return 0;
            }
        }
        return 1; 
    }
\end{lstlisting}

To illustrate the issue, consider a case where the secret password \textit{pwd} is the string \textit{"secret"}.
If the user input \textit{in} is also \textit{"secret"}, the comparison executes all six character checks by running the  $for$ loop $len$ times.
If the input is different from \textit{"secret"}, for example \textit{"secure"}, the comparison proceeds character by character.
In this case, the first three characters \textit{('s', 'e', 'c')} match, but the fourth comparison fails when \textit{'r'} is checked against \textit{'u'}.
At this point, the loop exits early, generating a shorter instruction trace compared to the earlier case.
The variability in the number of executed instructions makes the execution behavior dependent on the input, thereby violating the constant-time property of the program.

To prevent timing side-channel attacks in password verification, a constant-time implementation ensures that every character of the input is compared regardless of early mismatches. 
Listing~\ref{lst:cnst_password_checker} shows such an implementation. 
Instead of returning immediately upon a mismatch, it accumulates the \textit{XOR} differences of all characters and return the result at the end. 
This guarantees that the same sequence of instructions is executed for all inputs of the same length, eliminating input-dependent variations in execution thus constant-time.

\begin{lstlisting}[language=C, float, caption={Constant-time password comparison implementation}, label={lst:cnst_password_checker}]
    int cnst_password_checker(char *pw, char *in, int len) {
        int r = 0;
        for (int i = 0; i < len; i++) {
            r |= (pw[i] ^ in[i]);
        }
        return (r == 0); 
    }
\end{lstlisting}

We evaluated the same input values to this program as in the non-constant-time case above using \system. 
For the case with \textit{pw = "secret"}, traces were collected for inputs \textit{in1 = "secret"} and \textit{in2 = "secure"}. 
The analysis confirmed that the number and type of executed instructions remained identical across both inputs, unlike in the non-constant-time case. 

Table~\ref{tab:password_const_nonconst_cls} summarizes the classification of instructions collected by \system in these two cases and for the two sets of inputs discussed in this example. We show these results from binaries compiled with different compilers and optimizations.

\begin{table*}[ht]
\centering
\small
\caption{Instruction classification collected by \system for the password checking example with Input~1 and Input~2. This case study used 0 vector and floating point instructions in all cases so we removed those from the Table.}
\label{tab:password_const_nonconst_cls}
\resizebox{\textwidth}{!}{%
\begin{tabular}{|l|cc|cc|cc|cc|cc|cc|cc|cc|cc|cc|cc|cc|}
\hline
\multirow{3}{*}{\textbf{Instruction Type}} 
& \multicolumn{12}{c|}{\textbf{Constant-Time}} 
& \multicolumn{12}{c|}{\textbf{Non-Constant-Time}} \\
\cline{2-25}
& \multicolumn{6}{c|}{\textbf{GCC}} 
& \multicolumn{6}{c|}{\textbf{Clang}} 
& \multicolumn{6}{c|}{\textbf{GCC}} 
& \multicolumn{6}{c|}{\textbf{Clang}} \\
\cline{2-25}
 & \multicolumn{2}{c|}{\textbf{-O0}} 
 & \multicolumn{2}{c|}{\textbf{-O1}} 
 & \multicolumn{2}{c|}{\textbf{-O2}} 
 & \multicolumn{2}{c|}{\textbf{-O0}} 
 & \multicolumn{2}{c|}{\textbf{-O1}} 
 & \multicolumn{2}{c|}{\textbf{-O2}} 
 & \multicolumn{2}{c|}{\textbf{-O0}} 
 & \multicolumn{2}{c|}{\textbf{-O1}} 
 & \multicolumn{2}{c|}{\textbf{-O2}} 
 & \multicolumn{2}{c|}{\textbf{-O0}} 
 & \multicolumn{2}{c|}{\textbf{-O1}} 
 & \multicolumn{2}{c|}{\textbf{-O2}} \\
\cline{2-25}
 & In1 & In2 & In1 & In2 & In1 & In2 & In1 & In2 & In1 & In2 & In1 & In2 
 & In1 & In2 & In1 & In2 & In1 & In2 & In1 & In2 & In1 & In2 & In1 & In2 \\
\hline
Load 
                    & 259 & 259 & 111 & 111  & 108 & 108  & 257 & 257 & 108 & 108  & 115 & 115 
                    & 242 & 153 & 98 & 94 & 98 & 62 & 246 & 147 & 98 & 14 & 98 & 46 \\
\hline
Store & 0 & 0 & 0 & 0 & 0 & 0 & 0 & 0 & 0 & 0 & 0 & 0 
                    & 0 & 0 & 0 & 0  & 0 & 0  & 0 & 0 & 0 & 0  & 0 & 0  \\
\hline
Light ALU         
                    & 65 & 65 & 13  & 13  & 13  & 13  & 63 & 63 & 13 & 13  & 13 & 13  
                    & 58 & 37 & 13 & 13 & 13 & 9 & 57 & 36 & 13 & 3 & 13 & 7 \\
                    \hline
Heavy ALU         
                    & 12 & 12 & 0  & 0  & 0  & 0  & 12 & 12 & 0 & 0  & 0 & 0 
                    & 6 & 3 & 0 & 0 & 0 & 0 & 7 & 4 & 0 & 0 & 0 & 0 \\
\hline
Control-Flow 
                    & 8  & 8   & 10  & 10  & 10  & 10   & 8  & 8  & 10   & 10   & 10   & 10   
                    & 13  & 9   & 18  & 18  & 13  & 8  & 20   & 11   & 18  & 3  & 18  & 9   \\
\hline
\end{tabular}}
\end{table*}

\subsection{Modular Exponentiation Example}

Modular exponentiation is a fundamental operation in many cryptographic algorithms, including RSA. 
However, naive implementations generally exhibit non-constant-time behavior, making them vulnerable to timing side-channel attacks. 
Consider the naive implementation shown in Listing~\ref{lst:modexp_nonconst}:

\lstset{style=mystyle}
\begin{lstlisting}[language=C, float, caption=Naive modular exponentiation Example, label={lst:modexp_nonconst}]
    uint64_t modexp_nonconst(uint64_t base, uint64_t exp, uint64_t mod) {
        uint64_t result = 1;
        for (int i = 63; i >= 0; i--) {
            if ((exp >> i) & 1) { 
                result = (result * base) % mod;
            }
        }
        return result; 
    }
\end{lstlisting}

In this implementation, the loop iterates over all 64 bits of the secret exponent \textit{exp}. 
The multiplication and modular reduction operations inside the conditional statement are executed only when the current bit of \textit{exp} is \textit{1}. 
As a result, the number of executed instructions, memory accesses, and the overall control-flow path depend directly on the secret input.
For illustration, consider two 8-bit exponents: \textit{exp1 = 0b10001100 (decimal 140)} and \textit{exp2 = 0b11110000 (decimal 240)}. 
The first exponent \textit{(exp1)} contains three 1-bits, triggering multiplications in three iterations of the loop, while the second exponent \textit{(exp2)} contains 1-bits in the first digits iterations and zeros for the last four, triggering multiplications in the first four iterations. 
Instruction traces collected and analyzed with \system showed differences in executed instructions between these two cases. 

To mitigate the timing side-channel vulnerability in modular exponentiation, a constant-time implementation ensures that the same sequence of operations is executed for every bit of the exponent, regardless of its value. 
Listing~\ref{lst:modexp_const} shows such an implementation, where the squaring operation is always performed, and the multiplication is computed in a constant-time manner using bit masking. 
The mask ensures that the result is conditionally updated without introducing branches dependent on secret bits, eliminating input-dependent variations in the instruction sequence generated from the program.  
\begin{lstlisting}[language=C, float, caption={Constant-time modular exponentiation Example}, label={lst:modexp_const}]
    uint64_t modexp_const(uint64_t base, uint64_t exp, uint64_t mod) {
        uint64_t result = 1;
        for (int i = 63; i >= 0; i--) {
            result = (result * result) % mod;
            uint64_t mask = -((exp >> i) & 1ULL);
            uint64_t tmp = (result * base) % mod;
            result = (tmp & mask) | (result & ~mask);
        }
        return result; 
    }
\end{lstlisting}

We evaluated this constant-time implementation using \system by collecting instruction traces for multiple exponent values, including \textit{exp1 = 10001100 (140)} and \textit{exp2 = 0b11110000 (240)}. 
The analysis confirmed that, unlike the non-constant-time version, the number and types of executed instructions remained identical across all inputs. 
Table~\ref{tab:modexp_cls} summarizes the instruction classification collected for both the constant-time and non-constant-time modular exponentiation implementations, showing consistent instruction traces in the constant-time case and input-dependent variations in the non-constant-time example in all compilation pipelines.

\begin{table*}[ht]
\centering
\small
\caption{Instruction classification collected by \system for modulus exponentiation example with Input~1 and Input~2. This case study used 0 vector and floating point instructions in all cases so we removed those from the Table}
\label{tab:modexp_cls}
\resizebox{\textwidth}{!}{%
\begin{tabular}{|l|cc|cc|cc|cc|cc|cc|cc|cc|cc|cc|cc|cc|}
\hline
\multirow{3}{*}{\textbf{Instruction Type}} 
& \multicolumn{12}{c|}{\textbf{Constant-Time}} 
& \multicolumn{12}{c|}{\textbf{Non-Constant-Time}} \\
\cline{2-25}
& \multicolumn{6}{c|}{\textbf{GCC}} 
& \multicolumn{6}{c|}{\textbf{Clang}} 
& \multicolumn{6}{c|}{\textbf{GCC}} 
& \multicolumn{6}{c|}{\textbf{Clang}} \\
\cline{2-25}
 & \multicolumn{2}{c|}{\textbf{-O0}} 
 & \multicolumn{2}{c|}{\textbf{-O1}} 
 & \multicolumn{2}{c|}{\textbf{-O2}} 
 & \multicolumn{2}{c|}{\textbf{-O0}} 
 & \multicolumn{2}{c|}{\textbf{-O1}} 
 & \multicolumn{2}{c|}{\textbf{-O2}} 
 & \multicolumn{2}{c|}{\textbf{-O0}} 
 & \multicolumn{2}{c|}{\textbf{-O1}} 
 & \multicolumn{2}{c|}{\textbf{-O2}} 
 & \multicolumn{2}{c|}{\textbf{-O0}} 
 & \multicolumn{2}{c|}{\textbf{-O1}} 
 & \multicolumn{2}{c|}{\textbf{-O2}} \\
\cline{2-25}
 & In1 & In2 & In1 & In2 & In1 & In2 & In1 & In2 & In1 & In2 & In1 & In2 
 & In1 & In2 & In1 & In2 & In1 & In2 & In1 & In2 & In1 & In2 & In1 & In2 \\
\hline
Load 
                    & 2762 & 2762 & 1703 & 1703 & 1640 & 1640 & 2948 & 2948 & 1703 & 1703 & 3074 & 3074
                    & 1434 & 1442 & 1738 & 1734 & 1657 & 1661 & 1618 & 1611 & 1224 & 1227 & 1278 & 1292 \\
\hline
Store 
                    & 128 & 128 & 126 & 126 & 126 & 126 & 128 & 128 & 126 & 126 & 128 & 128 
                    & 2 & 3 & 2 & 1 & 2 & 3 & 2 & 1 & 2 & 3 & 2 & 3 \\
\hline
Light ALU           
                    & 836 & 836 & 1 & 1 & 1 & 1 & 899 & 899 & 64 & 64 & 65 & 65 
                    & 266 & 269 & 65 & 65 & 64 & 64 & 265 & 262 & 65 & 65 & 65 & 65 \\
                    \hline
Heavy ALU         
                    & 320 & 320 & 0 & 0 & 0 & 0 & 320 & 320 & 63 & 63 & 64 & 64
                    & 66 & 67 & 64 & 64 & 63 & 63 & 66 & 65 & 64 & 64 & 64 & 64 \\ 
\hline
Control-Flow 
                    & 130 & 130 & 126 & 126 & 126 & 126 & 65 & 65 & 64 & 64 & 194 & 194 
                    & 194 & 194 & 191 & 192 & 252 & 251 & 193 & 193 & 192 & 191 & 161 & 161 \\
\hline
\end{tabular}}
\end{table*}

To provide a comprehensive evaluation of \system, we shall make all our the other examples  publicly available as part   \system tool. These examples include a variety of cryptographic primitives, and arithmetic routines that are commonly used in security-critical applications. 
The results show that \system accurately distinguished between constant-time and non-constant-time programs in all cases. 
In particular, non-constant-time code always exhibited input-dependent variations in instruction sequences, while constant-time implementations maintained identical instruction traces independednt of the secret input values.

\section{Discussion}

Our approach to constant-time verification fundamentally differs from prior works that rely on source-level analysis or statistical timing measurements. 
By analyzing low-level instruction traces across multiple inputs, \system captures the concrete logical execution behavior of programs. 
This enables direct observation of input-dependent variations in instruction sequences, which are the primary source of timing side channels. 
Unlike prior approaches, our method reflects the actual logical program behavior during execution.
Direct comparisons with existing tools are challenging because many focus on formal verification at the source code level or timing-based statistical tests, which are fundamentally different in methodology and scope. 
Our evaluation demonstrates the effectiveness of \system across a wide range of examples, but its contributions should be viewed as complementary or as an alternative approach for verifying constant-time programs. 

One of the key advantages of \system is its efficiency as analyzing instruction traces is fast and lightweight, making it practical for evaluating large codebases.
However, this dynamic approach has a limitation as it can only detect constant-time violations for the inputs that are actually tested through the binary. 
Comprehensive verification therefore requires executing the binary over a  diverse input sets to cover variations of inputs in secret-dependent behavior. 
Integrating fuzzing or symbolic execution addresses this limitation. 
In our current implementation, we use a very simple fuzzer which simply mutate the inputs provided using the havoc fuzzing approach. 
An intelligent fuzzing engine can systematically generate diverse test inputs to explore different program paths, while a symbolic execution engine can reason about all feasible paths with respect to a secret input and ensure that only relevant test input values are generated~ \cite{zhu2022fuzzing},\cite{king1976symbolic}. 
Combining \system with these optimized techniques will provide a more efficient and scalable coverage of programs as well as very strong guarantees of verifying their constant-time properties.

It is also worth addressing that our method does not account for microarchitectural side channels. Even in identical traces, runtime may still vary due to microarchitectural state. For example, the retired cycles of memory instructions is influenced by whether the data was already present in the cache. The event of such variation creates an exploitable side channel for attacks such as prime and probe, and flush and reload~\cite{su2021survey}. While it may be possible to detect these by modeling cache behavior, pure simulation cannot accurately represent the complexity of modern CPUs, especially coupled with concepts such as prefetching, shared caches and operating system scheduling. Alternatively, our approach can be coupled with statistical methods~\cite{reparaz2017dude}. While these methods alone only confirm the presence of constant time violations and not their absence, \system can corroborate such methods, leading to improved confidence in findings. 
Nevertheless, our approach remains effective and sufficient for verifying all logical constant-time behavior.

\section{Conclusion}

Timing side-channel attacks remain a significant threat to cryptographic and other security-critical software, as even small variations in execution time can leak sensitive information. 
In this work, we propose a novel approach to constant-time verification based on low-level instruction trace analysis, which captures the concrete execution behavior of programs across multiple inputs irrespective of system software, compilers, optimization levels, and hardware behavior. 
Building on this approach, we develop a tool called \system that automatically analyzes low-level instruction traces to reliably detect deviations from constant-time behavior in programs. 

We evaluate \system on a wide range of program examples, including password comparison routines and modular exponentiation, as well as additional well-known constant-time and non-constant-time implementations. 
The results unanimously demonstrate that \system accurately distinguishes between constant-time and non-constant-time code in all cases, providing a practical and lightweight method for detecting potential timing attacks in programs. 
While dynamic trace analysis is efficient, achieving comprehensive verification requires diverse input sets, which can be achieved by integrating fuzzing or symbolic execution engines. 
Overall, \system provides a robust and practical solution for developers seeking to secure software against timing side-channel attacks in the real-world by effectively analyzing low-level instructions of any program.

\bibliographystyle{plain}
\bibliography{paper}

\end{document}